\documentclass[sigconf]{acmart}

\usepackage{graphicx}
\usepackage{caption}
\usepackage{subcaption}
\usepackage{booktabs} 
\usepackage{xcolor, colortbl}
\usepackage{siunitx}
\usepackage[utf8]{inputenc}
\usepackage[moderate]{savetrees}

\setcopyright{rightsretained}

\begin{document}
\fancyhead{}

\title{Seasonal Web Search Query Selection for Influenza-Like Illness \\ (ILI) Estimation}

\author{Niels Dalum Hansen}
\affiliation{%
  \institution{University of Copenhagen / IBM Denmark}
}
\email{nhansen@di.ku.dk}

\author{Kåre Mølbak}
\affiliation{%
  \institution{Statens Serum Institut, Denmark}
}
\email{krm@ssi.dk}

\author{Ingemar J. Cox}
\affiliation{%
  \institution{University of Copenhagen, Denmark}
}
\email{ingemar.cox@di.ku.dk}

\author{Christina Lioma}
\affiliation{%
    \institution{University of Copenhagen, Denmark}
}
\email{c.lioma@di.ku.dk}

\begin{abstract}
\textit{Influenza-like illness} (ILI) estimation from web search data is an important web analytics task. The basic idea is to use the frequencies of queries in web search logs that are correlated with past ILI activity as features when estimating current ILI activity. It has been noted that since influenza is seasonal, this approach can lead to spurious correlations with features/queries that also exhibit seasonality, but have no relationship with ILI. Spurious correlations can, in turn, degrade performance. To address this issue, we propose modeling the seasonal variation in ILI activity and selecting queries that are correlated with the residual of the seasonal model and the observed ILI signal. Experimental results show that re-ranking queries obtained by Google Correlate based on their correlation with the residual strongly favours ILI-related queries. 


\end{abstract}

%
%


\copyrightyear{2017} 
\acmYear{2017} 
\setcopyright{acmcopyright}
\acmConference{SIGIR’17}{}{August 7-11, 2017, Shinjuku, Tokyo, Japan}
\acmPrice{15.00}\acmDOI{http://dx.doi.org/10.1145/3077136.3080760}
\acmISBN{978-1-4503-5022-8/17/08}

\maketitle

\section{Introduction and background}

The frequency of queries in web search logs has been found useful in estimating the incidence of \textit{influenza-like illnesses} (ILIs) \cite{ginsberg2009detecting,lampos2015advances,lazer2014parable,santillana2015combining,yang2015accurate}. 
Current methods use two core discriminative features for ILI estimation: (i) past ILI activity, and (ii) the frequency of queries in web search logs that correlate strongly with past ILI activity.  There are two problems with this approach.

The \textit{first problem} is that not all queries whose frequency is strongly correlated to ILI activity are necessarily informative with respect to ILIs, and hence discriminative as a feature for ILI estimation.
At the most general level, this is an example of the fundamental issue of correlation not being causation. In the case of estimating ILI, this is exacerbated by the seasonal nature of influenza. In fact, it has been previously observed that previous methods can identify queries that have a very similar seasonality but are clearly not related to ILI. For example, the query \textit{``high school basketball''} has been found to have a high correlation with ILI activity [8] even though it is obviously unrelated to ILI. The seasonality of high school basketball accounts for this correlation. 
Queries unrelated to the ILI activity will not be useful in the case of irregular ILI activity, e.g. an off season influenza outbreak. Additionally, changes in for example the high school basketball schedule would result in changes in the ILI estimates. 

The \textit{second problem} is that by using two types of features that are strongly correlated to each other (past ILI activity, and queries whose frequencies are strongly correlated to past ILI activity), we may compromise diversity in the representations one would expect from the features. Better estimations may be produced by using features that \textit{complement} each other, regardless of their between--feature correlation. 

Motivated by the above issues, we 
propose an alternative approach to selecting queries.
Our approach consists of two steps. (1) We model the seasonal variation of ILI activity, and (2) we select queries whose search frequency fits aspects of this seasonality. Specifically, we present two variations of our algorithm: select queries that 
correlate with our seasonal model of ILI,
and select queries that correlate with the residual between the seasonal model and observed ILI rates.

Our results are two fold. (i) Experimental evaluation of our seasonal query selection models for ILI estimation against strong recent baselines (of no seasonality) show that we can achieve performance that is overall more effective (reduced estimation error), and requires fewer queries as estimation features. With respect to error reduction we see that selecting queries fitting regular seasonal ILI variation is a better strategy than selecting queries fitting ILI outbreaks. (ii) Selecting queries that fit seasonal irregularities result in much more semantically relevant queries. These queries are surprisingly not the ones that result in the best predictions. 

Our main results are: (i) We demonstrate that Google Correlate retrieves many non-relevant queries that are highly correlated with a times series of historic ILI incidence, and that the ILI-related queries are not highly ranked; (ii) re-ranking these queries based on their correlation with a residual signal, i.e. the difference between a seasonal model and historic data, strongly favours ILI-related queries; (iii) the performance of a linear estimator is improved based on the re-ranked queries. 
To our knowledge, the seasonal examination of ILI activity for query selection in automatic ILI estimation is a novel contribution. Seasonal variation has, however, been studied for other medical informatics tasks, such as vaccination uptake estimation \cite{www17, HansenLM16}.


\section{Problem Statement}
The goal is to estimate ILI activity at time $t$, denoted $y_t$, using observed historical ILI activity (reported e.g. by the Centers for Disease Control and Prevention (CDC) in US) and query frequencies in web search logs. This is most commonly done by (i) submitting to Google Correlate\footnote{\url{https://www.google.com/trends/correlate}} a file of historical ILI activity, and receiving as output web search queries (and their frequencies) that are most strongly correlated to the input ILI data. Then, $y_t$ can be estimated with a linear model that uses only web search frequencies \cite{ginsberg2009detecting} as follows:

\begin{equation}\label{eq:queries}
y_t = \alpha_0 + \sum^n_{i=1} \alpha_i Q_{t,i} + \epsilon,
\end{equation}

\noindent where $n$ is the number of queries, $Q_{t,i}$ is the frequency for query $i$ in the web search log at time $t$, the $\alpha$s are coefficients, and $\epsilon$ is the estimation error. 

Including historical ILI activity data can improve the estimations of Eq. \ref{eq:queries}, for instance with an autoregressive model \cite{yang2015accurate}, as follows:

\begin{equation}\label{eq:clinical+queries}
y_t = \beta_0 + \beta_1 t + \sum^m_{j=1} \beta_{j+1} \cdot y_{t-j} + \sum^n_{i=1} \beta_{i+m+1} \cdot Q_{t,i}+ \epsilon, 
\end{equation}

\noindent where $m$ is the number of autoregressive terms, and the $\beta$s are coefficients to be learned.
With $m=52$ and $n=100$, Eq. \ref{eq:clinical+queries} corresponds to the model presented by Yang {\em et al.}~\cite{yang2015accurate}. 

Most ILI estimation methods (exceptions include \cite{DBLP:conf/www/LamposZC17}) that use web search frequencies use all queries found to be correlated to ILI activity, i.e. in Eq. \ref{eq:queries} and Eq. \ref{eq:clinical+queries} $n$ corresponds to \textit{all} strongly correlated queries, and query selection is typically left for the model regularisation, for example using lasso regularisation. In the next section we present a novel way of selecting which among these correlated queries to include in the estimation of $y_t$ according to how well they fit the seasonal variation of ILI activity. 

\section{Seasonal Query Selection}\label{sec:seasonal_query_selection}

We reason that among the queries whose frequency is correlated with past ILI activity, some queries may fit the ILI seasonal variation better than others. This is supported by the literature \cite{lazer2014parable}. We further reason that this fit of queries to seasonal ILI variation may not be sufficiently captured by simply measuring the correlation between the frequency of those queries and ILI activity. Based on this, we (i) present two  models to represent seasonal variation of ILI activity, and (ii) select queries 
based on these seasonal models.

\subsection{Step 1: Model seasonal ILI variation}

\begin{figure}
\centering
\scalebox{.5}{
\includegraphics[]{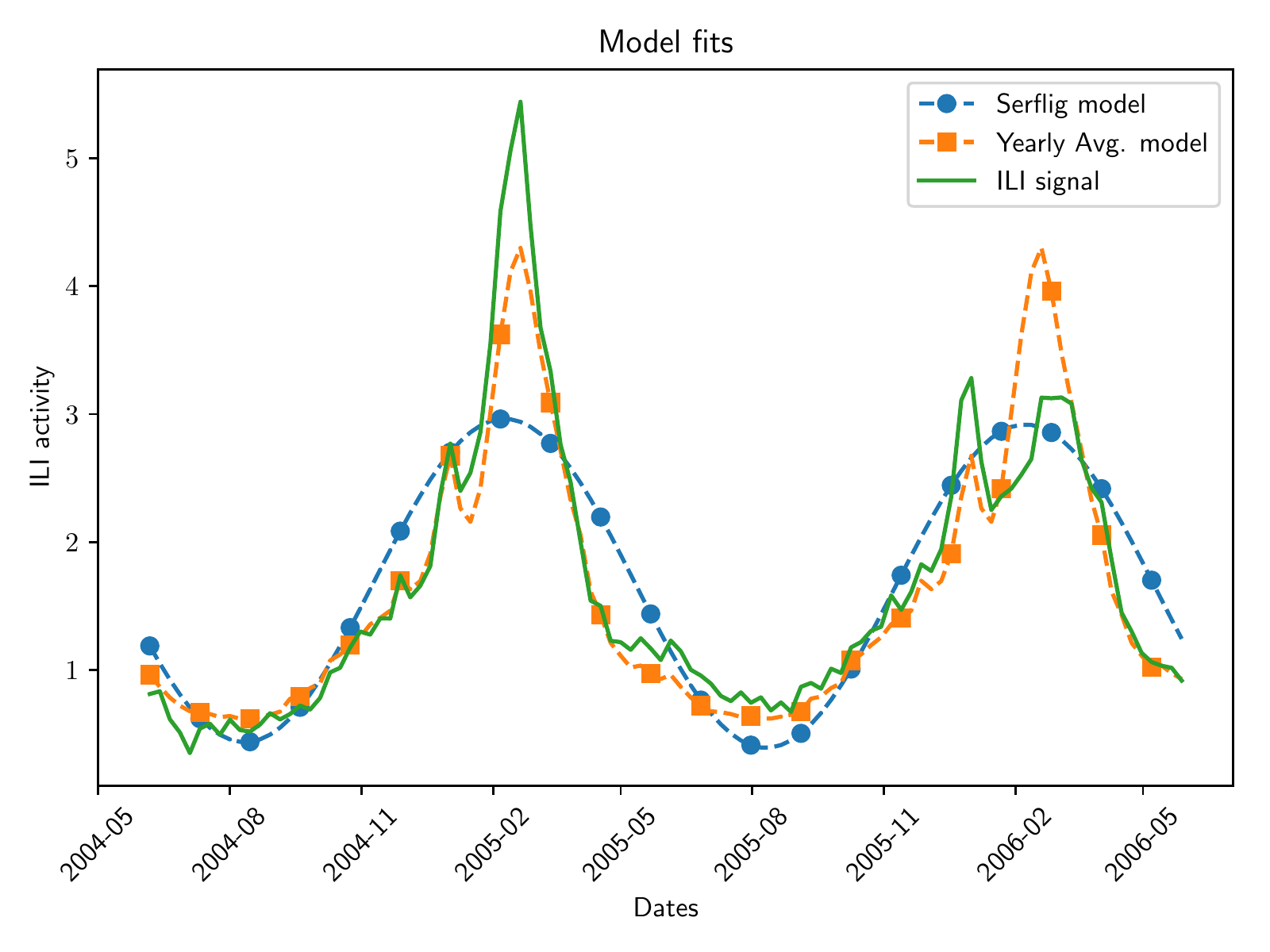}
}
\caption{Fit of the Serfling model (Eq. \ref{eq:serf}) and the Yearly Average model (Eq. \ref{eq:avg}) to historical ILI data (described in Section \ref{s:eval}).}
\label{fig:serflingfit}
\end{figure}


We model seasonal variation in two ways.
The first model is the Serfling model \cite{serfling1963methods}, chosen because of its simplicity and expressiveness. The Serfling model (Eq. \ref{eq:serf}) uses pairs of sine and cosine terms to model seasonality, and a term linear in time to model general upward or downward trends. We use this model with weekly data and one yearly cycle (details on data are given in Section \ref{s:eval}), resulting in the following ILI estimation model:

\begin{equation}\label{eq:serf}
y_t = \beta_0 + \beta_1 t + \beta_2 \sin\left(\frac{2 \pi t}{52}\right) + \beta_3 \cos\left(\frac{2 \pi t}{52}\right) + \epsilon,
\end{equation}

\noindent where the $\beta$s denote model coefficients and $\epsilon$ the error, i.e. residual. 

For the second model we use a yearly average (YA). Here the expected value of $y_t$ is calculated as the average value of $N$ seasons of ILI activity data, 

\begin{equation}\label{eq:avg}
\hat{y}_t = \frac{1}{N} \sum_{i=0}^{N-1} y_{(t \text{ mod } S) + i \cdot S},
\end{equation}

\noindent where $S$ is the season length in weeks, in our case 52.

Fig. \ref{fig:serflingfit} shows the fit of the two models, i.e. the Serfling model (Eq. \ref{eq:serf}) and the YA model (Eq. \ref{eq:avg}) to historical ILI activity data. We see that the Serfling model fits the general seasonality, but not more complex patterns representing higher transmission. It does not, for example, model differences between the start and end of the ILI season. This is better captured by the YA model. 


\subsection{Step 2: Query selection}\label{ss:qs}

Having modelled seasonal ILI variation, the second step is to approximate how well queries fit the seasonal variation of ILI activities modelled by Eq. \ref{eq:serf}-\ref{eq:avg}. We do this in two ways:

\textbf{Seasonal correlation.} We compute the Pearson correlation between the query frequencies and the ILI seasonal model, i.e. Eq. \ref{eq:serf} or \ref{eq:avg}. We then select queries that are most strongly correlated to the ILI activity model.

\textbf{Residual correlation.} We compute the Pearson correlation between the query frequencies and the residual between the ILI seasonal model and the historical ILI activity. We then select queries that are most strongly correlated to the residual, i.e. \textit{unexpected variations in ILI activity} (possible outbreaks). 

The four query selection methods are denoted (i) Seasonal (Serfling), (ii) Seasonal (YA), (iii) Residual (Serfling), and (iv) Residual (YA).

\section{Evaluation}
\label{s:eval}

\paragraph{\textbf{Experimental Setup}}
We experimentally evaluate our seasonality-based query selection methods using two types of data: weekly ILI activity data and Google search frequency data. The ILI activity data is from the US CDC for the period 2004-6-6 to 2015-7-11\footnote{\url{https://gis.cdc.gov/grasp/fluview/fluportaldashboard.html}} (inclusive). The CDC reports this in the form of weighted ILI activity across different US regions. ILI activity for a region corresponds to the number of ILI related visits to health care providers compared to non-influenza weeks, e.g. an ILI activity of 2 corresponds to twice as many visits as in non-epidemic weeks.

We retrieve query search frequencies from Google Correlate with the location set to the US. Specifically, we use the 100\footnote{This is a maximum number set by Google Correlate.} queries that have the highest correlation with the ILI activity from 2004-6-6 to 2009-3-29 according to Google Correlate. Google normalizes the search frequencies for each query to unit variance and zero mean, i.e. we do not know the actual search frequencies. We use the interval 2004-6-6 to 2009-3-29 because it represents a non-epidemic period (it excludes the 2009 pandemic of H1N1 influenza virus that caused highly irregular ILI activity). The 100 queries are shown in Tab.~\ref{tabel:ili_queries}. Only 21 of the 100 queries are related to ILI (in bold).

We compare our query selection methods (Section \ref{sec:seasonal_query_selection}) to the following three baselines: (Tab. \ref{tab:resultsQuery} baseline i) uses the top-$c$ queries to estimate ILI activity, where the top-$c$ are chosen to minimise the RMSE error, i.e. if we use $c+1$ queries the RMSE increases; (Tab. \ref{tab:resultsQuery} baseline ii) using \textbf{\textit{all}} 100 queries to estimate ILI activity; (Tab. \ref{tab:resultsQuery} baseline iii) using no queries, only past ILI activity, i.e. an autoregressive model.  For (i) and (ii), the query ranking is determined by Google Correlate. For (iii), two autoregressive models are fitted: one using 3 autoregressive terms \cite{lazer2014parable,yang2015accurate} and one with 52 autoregressive terms \cite{yang2015accurate}. This setup is similar to the setup of Yang {\em et al.}~\cite{yang2015accurate}. We implement baseline (iii) using Eq. \ref{eq:clinical+queries} where $m$ is set to 3 and 52 terms, respectively, and $n=0$. Similarly to \cite{lazer2014parable,yang2015accurate}, we evaluate estimation performance by reporting the root mean squared error (RMSE) and Pearson correlation between the estimations and the observed historical ILI activity.

For all runs, we use data from 2004-6-1 to 2009-3-29 for training, and data from 2009-4-1 to 2015-7-11 for testing. The training data is used to fit Eq. \ref{eq:serf}-\ref{eq:avg}, and to calculate the correlation scores as described in Section \ref{ss:qs}. Estimations are made in a leave-one-out fashion where data prior to the data point being estimated is used to fit the estimation model. Each model is retrained for every time step using the 104 most recent data points (exactly as in Yang {\em et al.}~\cite{yang2015accurate}). We determine the number of queries $n$ in Eq. \ref{eq:queries}-\ref{eq:serf} by iteratively adding the next highest ranked query, where query rank is given by either Google Correlate (for the baselines), or by the four variants of our algorithm, specifically (i) correlate seasonal (Serfling), (ii) correlate seasonal (YA), (iii) correlate residual (Serfling), and (iv) correlate residual (YA). The models are fitted using lasso regression, where the hyper-parameter is found using three fold cross-validation on the training set.

\paragraph{\textbf{Results}}

\begin{table*}
	\begin{tabular}{p{16cm}}
		\toprule
		\footnotesize{
		florida spring, \textbf{influenza symptoms}, \textbf{symptoms of influenza}, new york yankees spring training, yankees spring training, \textbf{flu incubation}, \textbf{flu incubation period}, \textbf{flu duration}, florida spring training, spring training locations, \textbf{influenza incubation}, florida spring break, spring break dates, \textbf{flu fever}, sox spring training, new york mets spring training, \textbf{bronchitis}, red sox spring, spring training in florida, snow goose, spring break calendar, spring training in arizona, red sox spring training, mlb spring training, \textbf{flu report}, baseball spring training, mariners spring training, wrestling tournaments, spring training, golf dome, \textbf{flu recovery}, city spring, wrestling singlets, spring training sites, boys basketball, \textbf{type a influenza}, yankee spring training, spring training tickets, las vegas march, indoor track, harlem globe, spring break panama city, girls basketball, panama city spring break, cardinals spring training, ny mets spring training, ny yankees spring training, \textbf{flu symptoms}, minnesota twins spring training, concerts in march, spring training map, \textbf{tessalon}, boston red sox spring training, \textbf{flu contagious}, \textbf{symptoms of the flu}, events in march, seattle mariners spring training, singlets, \textbf{influenza contagious}, \textbf{influenza incubation period}, spring break schedule, spring vacation, \textbf{treating the flu}, college spring break, basketball boys, college spring break dates, boys basketball team, \textbf{respiratory flu}, atlanta braves spring training, \textbf{acute bronchitis}, march madness dates, spring break florida, braves spring training, college basketball standings, in march, braves spring training schedule, high school boys basketball, spring break ideas, spring break miami, banff film, addy awards, grapefruit league, spring clothing, spring collection, banff film festival, st. louis cardinals spring training, april weather, spring break family, red sox spring training schedule, miami spring break, nj wrestling, spring break getaways, spring break date, high school boys, march concerts, high school basketball, indoor track results, \textbf{tussionex}, globetrotters, orioles spring training}\\
		\bottomrule
	\end{tabular}
	\caption{The 100 queries retrieved from Google Correlate. We treat queries in bold as ILI related.}
	\label{tabel:ili_queries}
\end{table*}

\begin{figure}
	\centering
	\includegraphics[width=8.5cm]{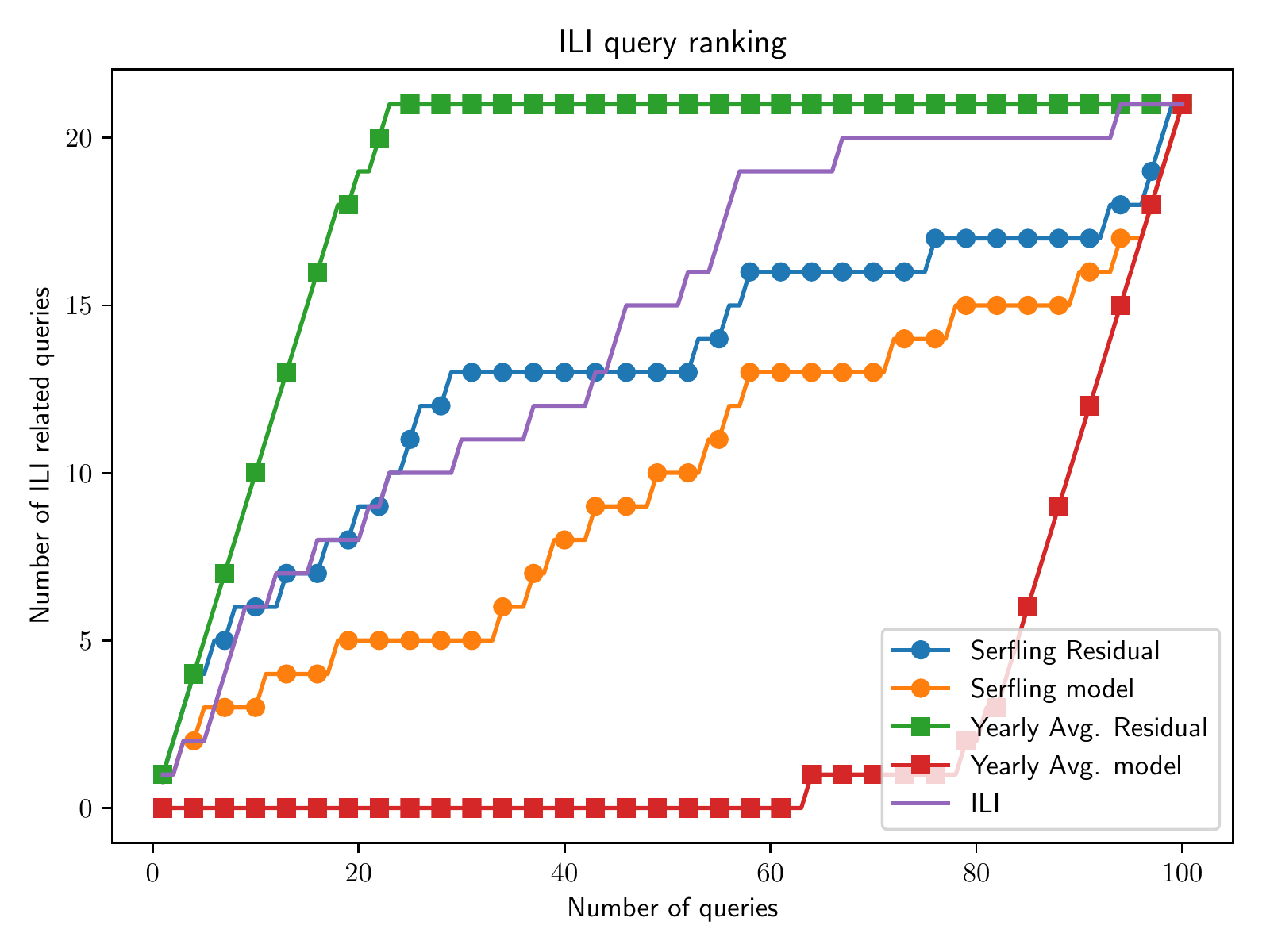}
	\caption{Portion of ILI related queries in the top $n$ queries for each of the five ranking methods.}
	\label{fig:ILIrelevant}
\end{figure}

As noted earlier, Google Correlate identifies the top-100 queries, but only 21 of these are ILI-related. Our four algorithms re-rank the 100 queries. Fig.~\ref{fig:ILIrelevant} plots the number of ILI-related queries as a function of the number of rank-ordered queries. The solid curve is based on the original ranking provided by Google Correlate. We observe that both the (i) Seasonal (Serfling) and (ii) Seasonal (YA), re-rank the queries such that, in general, the ILI-related queries are ranked worse. The Residual (Serfling) generally performs similarly to or worse than Google Correlate in favouring ILI-related queries. In contrast, Residual (YA) re-ranks the queries such that almost all ILI-related queries are favoured. Of the top-21 queries, 19 are ILI-related. All 21 ILI-related queries are within the top-23. The only two non-related queries in the top-23 are ranked at 19 and 21.  Clearly re-ranking queries based on Residual (YA) strongly favours ILI-related queries much more than Google Correlate or our other three variants.

For each ranking of the queries, we select the top-$n$ queries that either minimise the RMSE or maximise the Pearson correlation. This is done for the Linear model of Eq.~\ref{eq:queries} and for the autoregressive model of Eq.~\ref{eq:clinical+queries},
Tab. \ref{tab:resultsQuery} shows the results.
For the Linear model (column 1), we observe that Residual (YA) performs best w.r.t. RMSE and Pearson correlation, though the latter is not significant. Note that in both cases, (i) the number of queries needed by Residual (YA) is significantly less than for the other three variants and (ii) the two baselines performed worse.

For the autoregressive models, we observe that the Seasonal (Serfling) model performs best w.r.t. RMSE and Pearson correlation. This is achieved with relatively few queries (5, 9, or 11). However we note that of the top-5, -9 or -11 queries only 3, 3 or 4, resp. are ILI-related. In general, autoregressive models perform well when the signal has a strong autocorrelation. However, should the signal strongly deviate from seasonal patterns, then it is unlikely that the ILI estimates would be accurate.   


\begin{table*}
{
	\begin{tabular}{l l rr rr rr}
	
	&  & \multicolumn{6}{c}{\textbf{RMSE} \textit{(the lower, the better)}} \\
\toprule
	& & Linear (Eq. \ref{eq:queries}) & \#q. & Autoregressive 52 (Eq. \ref{eq:clinical+queries}) & \#q. & Autoregressive 3 (Eq. \ref{eq:clinical+queries}) & \#q. \\
 \cmidrule(lr){3-4} \cmidrule(lr){5-6} \cmidrule(lr){7-8}
Seasonal (Serfling)      &  & 0.398 & 97 & \textbf{0.280}& 5 & \textbf{0.280} & 11 \\
Residual (Serfling)  &  & 0.407 & 98 & 0.309 & 98 & 0.312 & 98 \\
Seasonal (YA)     &  & 0.394 & 96 & 0.311 & 100 & 0.298 & 68 \\
Residual (YA)  &  & 0.390 & 79 & 0.309 & 49 & 0.310 & 47 \\
Not Seasonal        & \textit{baseline i} & 0.413 & 33 & 0.309 & 68 & 0.312 & 46\\
Not Seasonal all q. & \textit{baseline ii} & 0.416 &    & 0.310 &    & 0.314 &\\
ILI History         & \textit{baseline iii} & n/a   &    & 0.348 &    & 0.333 &\\
\\
	&  & \multicolumn{6}{c}{\textbf{Correlation} \textit{(the higher, the better)}} \\
\toprule
& & Linear (Eq. \ref{eq:queries}) & \#q. & Autoregressive 52 (Eq. \ref{eq:clinical+queries}) & \#q. & Autoregressive 3 (Eq. \ref{eq:clinical+queries}) & \#q. \\
\cmidrule(lr){3-4} \cmidrule(lr){5-6} \cmidrule(lr){7-8}
Seasonal (Serfling)      &  & 0.948 & 96 & 0.973  & 5 & \textbf{0.974} & 9 \\
Residual (Serfling)   &  & 0.946  & 98 &0.968 & 100 &  0.967  & 98 \\
Seasonal (YA)    &  & 0.948 & 99 & 0.969 & 99 & 0.971 & 68 \\
Residual (YA)  &  & 0.949 & 77 &  0.969 & 59 & 0.966 & 47 \\
Not Seasonal        & \textit{baseline i} & 0.942 & 86 & 0.968 & 68 & 0.967 & 46 \\
Not Seasonal all q. & \textit{baseline ii} & 0.941 &   & 0.967 &    & 0.967 & \\
ILI History         & \textit{baseline iii} & n/a   &  & 0.959  & & 0.962  &\\ 
\bottomrule
	\end{tabular}
	}
\caption{Root mean squared error (RMSE) and Pearson Correlation of our seasonal ILI estimation methods and the three baselines. Bold marks the best score. \#q denotes the number of queries used in the estimation.}
	\label{tab:resultsQuery}
\end{table*}

\section{Conclusion}

The incidence of influenza-like illness (ILI) exhibits strong seasonal variations. These seasonal variations cause Google Correlate to identify a large number of non-relevant queries (~80\%). Many of the relevant queries are not highly ranked. Estimating the incidence of ILI with non-relevant queries is likely to become problematic when ILI deviates significantly from its seasonal variation.

We proposed a new approach to ILI estimation using web search queries. The novelty of our approach consists of re-ranking queries derived from Google Correlate. We first developed two models of the seasonal variation in ILI. The first is an analytical Serfling model. The second is an empirical model based on yearly averages. Four methods of re-ranking queries were then examined. The first two re-rank the queries based on their correlation with the two seasonal models. The second two re-rank queries based on their correlation with the residual between the seasonal models and historical ILI activity.

Experimental results showed that re-ranking queries based on Residual (YA) strongly favoured ILI-related queries, but 
re-ranking queries based on the two seasonal models, Seasonal (Serfling) and Seasonal (YA) led to rankings that were worse than those of Google Correlate.

When ILI estimates were based on both queries and autoregression the best performance was obtained when queries were re-ranked based on Seasonal (Serfling). Future work is needed to determine why, but we reason that 
(i) autoregessive models perform better when the signal has strong autocorrelation, i.e. is strongly seasonal, and (ii) this strong seasonality was present in our dataset, i.e. there was little deviation from the seasonal models. If, however, strong deviations did arise, we expect that models based on autoregression and queries re-ranked based on correlation with seasonal models will perform much worse.

This work complements the use of information retrieval and machine learning methods in the wider area of medical and health informatics \cite{DragusinPLLJW11,DragusinPLLJCHIW13}.

\paragraph{Acknowledgments.} Partially supported by Denmark's Innovation Fund (grant no. 110976).


\begin{thebibliography}{00}

{\scriptsize
	
	
	\ifx \showCODEN    \undefined \def \showCODEN     #1{\unskip}     \fi
	\ifx \showDOI      \undefined \def \showDOI       #1{{\tt DOI:}\penalty0{#1}\ }
	\fi
	\ifx \showISBNx    \undefined \def \showISBNx     #1{\unskip}     \fi
	\ifx \showISBNxiii \undefined \def \showISBNxiii  #1{\unskip}     \fi
	\ifx \showISSN     \undefined \def \showISSN      #1{\unskip}     \fi
	\ifx \showLCCN     \undefined \def \showLCCN      #1{\unskip}     \fi
	\ifx \shownote     \undefined \def \shownote      #1{#1}          \fi
	\ifx \showarticletitle \undefined \def \showarticletitle #1{#1}   \fi
	\ifx \showURL      \undefined \def \showURL       #1{#1}          \fi
	\providecommand\bibfield[2]{#2}
	\providecommand\bibinfo[2]{#2}
	\providecommand\natexlab[1]{#1}
	\providecommand\showeprint[2][]{arXiv:#2}
	
	\bibitem[\protect\citeauthoryear{Dragusin, Petcu, Lioma, Larsen, J{\o}rgensen,
		Cox, Hansen, Ingwersen, and Winther}{Dragusin et~al\mbox{.}}{2013}]%
	{DragusinPLLJCHIW13}
	\bibfield{author}{\bibinfo{person}{Radu Dragusin}, \bibinfo{person}{Paula
			Petcu}, \bibinfo{person}{Christina Lioma}, \bibinfo{person}{Birger Larsen},
		\bibinfo{person}{Henrik J{\o}rgensen}, \bibinfo{person}{Ingemar~J. Cox},
		\bibinfo{person}{Lars~K. Hansen}, \bibinfo{person}{Peter Ingwersen}, {and}
		\bibinfo{person}{Ole Winther}.} \bibinfo{year}{2013}\natexlab{}.
	\newblock \showarticletitle{Specialised Tools are Needed when Searching the Web
		for Rare Disease Diagnoses}.
	\newblock \bibinfo{journal}{{\em Rare Diseases\/}} \bibinfo{volume}{1},
	\bibinfo{number}{2} (\bibinfo{year}{2013}), \bibinfo{pages}{e25001--1 --
		e25001--4}.
	\newblock
	\showDOI{%
		\url{http://dx.doi.org/10.4161/rdis.25001}}
	
	
	\bibitem[\protect\citeauthoryear{Dragusin, Petcu, Lioma, Larsen, J{\o}rgensen,
		and Winther}{Dragusin et~al\mbox{.}}{2011}]%
	{DragusinPLLJW11}
	\bibfield{author}{\bibinfo{person}{Radu Dragusin}, \bibinfo{person}{Paula
			Petcu}, \bibinfo{person}{Christina Lioma}, \bibinfo{person}{Birger Larsen},
		\bibinfo{person}{Henrik J{\o}rgensen}, {and} \bibinfo{person}{Ole Winther}.}
	\bibinfo{year}{2011}\natexlab{}.
	\newblock \showarticletitle{Rare Disease Diagnosis as an Information Retrieval
		Task}. In \bibinfo{booktitle}{{\em Advances in Information Retrieval Theory -
			Third International Conference, {ICTIR} 2011, Bertinoro, Italy, September
			12-14, 2011. Proceedings}} {\em (\bibinfo{series}{Lecture Notes in Computer
			Science})}, \bibfield{editor}{\bibinfo{person}{Giambattista Amati} {and}
		\bibinfo{person}{Fabio Crestani}} (Eds.), Vol.~\bibinfo{volume}{6931}.
	\bibinfo{publisher}{Springer}, \bibinfo{pages}{356--359}.
	\newblock
	\showISBNx{978-3-642-23317-3}
	\showDOI{%
		\url{http://dx.doi.org/10.1007/978-3-642-23318-0_38}}
	
	
	\bibitem[\protect\citeauthoryear{Ginsberg, Mohebbi, Patel, Brammer, Smolinski,
		and Brilliant}{Ginsberg et~al\mbox{.}}{2009}]%
	{ginsberg2009detecting}
	\bibfield{author}{\bibinfo{person}{Jeremy Ginsberg}, \bibinfo{person}{Matthew~H
			Mohebbi}, \bibinfo{person}{Rajan~S Patel}, \bibinfo{person}{Lynnette
			Brammer}, \bibinfo{person}{Mark~S Smolinski}, {and} \bibinfo{person}{Larry
			Brilliant}.} \bibinfo{year}{2009}\natexlab{}.
	\newblock \showarticletitle{Detecting influenza epidemics using search engine
		query data}.
	\newblock \bibinfo{journal}{{\em Nature\/}} \bibinfo{volume}{457},
	\bibinfo{number}{7232} (\bibinfo{year}{2009}), \bibinfo{pages}{1012--1014}.
	\newblock
	
	
	\bibitem[\protect\citeauthoryear{Hansen, Lioma, and M{\o}lbak}{Hansen
		et~al\mbox{.}}{2016}]%
	{HansenLM16}
	\bibfield{author}{\bibinfo{person}{Niels~Dalum Hansen},
		\bibinfo{person}{Christina Lioma}, {and} \bibinfo{person}{K{\aa}re
			M{\o}lbak}.} \bibinfo{year}{2016}\natexlab{}.
	\newblock \showarticletitle{Ensemble Learned Vaccination Uptake Prediction
		using Web Search Queries}. In \bibinfo{booktitle}{{\em Proceedings of the
			25th {ACM} International on Conference on Information and Knowledge
			Management, {CIKM} 2016, Indianapolis, IN, USA, October 24-28, 2016}},
	\bibfield{editor}{\bibinfo{person}{Snehasis Mukhopadhyay},
		\bibinfo{person}{ChengXiang Zhai}, \bibinfo{person}{Elisa Bertino},
		\bibinfo{person}{Fabio Crestani}, \bibinfo{person}{Javed Mostafa},
		\bibinfo{person}{Jie Tang}, \bibinfo{person}{Luo Si},
		\bibinfo{person}{Xiaofang Zhou}, \bibinfo{person}{Yi~Chang},
		\bibinfo{person}{Yunyao Li}, {and} \bibinfo{person}{Parikshit Sondhi}}
	(Eds.). \bibinfo{publisher}{{ACM}}, \bibinfo{pages}{1953--1956}.
	\newblock
	\showISBNx{978-1-4503-4073-1}
	\showDOI{%
		\url{http://dx.doi.org/10.1145/2983323.2983882}}
	
	
	\bibitem[\protect\citeauthoryear{Hansen, M{\o}lbak, Cox, and Lioma}{Hansen
		et~al\mbox{.}}{2017}]%
	{www17}
	\bibfield{author}{\bibinfo{person}{Niels~Dalum Hansen},
		\bibinfo{person}{K{\aa}re M{\o}lbak}, \bibinfo{person}{Ingemar~J. Cox}, {and}
		\bibinfo{person}{Christina Lioma}.} \bibinfo{year}{2017}\natexlab{}.
	\newblock \showarticletitle{Time-Series Adaptive Estimation of Vaccination
		Uptake Using Web Search Queries}. In \bibinfo{booktitle}{{\em Proceedings of
			the 26th International Conference on World Wide Web Companion, Perth,
			Australia, April 3-7, 2017}}, \bibfield{editor}{\bibinfo{person}{Rick
			Barrett}, \bibinfo{person}{Rick Cummings}, \bibinfo{person}{Eugene
			Agichtein}, {and} \bibinfo{person}{Evgeniy Gabrilovich}} (Eds.).
	\bibinfo{publisher}{{ACM}}, \bibinfo{pages}{773--774}.
	\newblock
	\showISBNx{978-1-4503-4914-7}
	\showDOI{%
		\url{http://dx.doi.org/10.1145/3041021.3054251}}
	
	
	\bibitem[\protect\citeauthoryear{Lampos, Miller, Crossan, and Stefansen}{Lampos
		et~al\mbox{.}}{2015}]%
	{lampos2015advances}
	\bibfield{author}{\bibinfo{person}{Vasileios Lampos}, \bibinfo{person}{Andrew~C
			Miller}, \bibinfo{person}{Steve Crossan}, {and} \bibinfo{person}{Christian
			Stefansen}.} \bibinfo{year}{2015}\natexlab{}.
	\newblock \showarticletitle{Advances in nowcasting influenza-like illness rates
		using search query logs}.
	\newblock \bibinfo{journal}{{\em Scientific reports\/}}  \bibinfo{volume}{5}
	(\bibinfo{year}{2015}).
	\newblock
	
	
	\bibitem[\protect\citeauthoryear{Lampos, Zou, and Cox}{Lampos
		et~al\mbox{.}}{2017}]%
	{DBLP:conf/www/LamposZC17}
	\bibfield{author}{\bibinfo{person}{Vasileios Lampos}, \bibinfo{person}{Bin
			Zou}, {and} \bibinfo{person}{Ingemar~Johansson Cox}.}
	\bibinfo{year}{2017}\natexlab{}.
	\newblock \showarticletitle{Enhancing Feature Selection Using Word Embeddings:
		The Case of Flu Surveillance}. In \bibinfo{booktitle}{{\em Proceedings of the
			26th International Conference on World Wide Web, {WWW} 2017, Perth,
			Australia, April 3-7, 2017}}. \bibinfo{pages}{695--704}.
	\newblock
	\showDOI{%
		\url{http://dx.doi.org/10.1145/3038912.3052622}}
	
	
	\bibitem[\protect\citeauthoryear{Lazer, Kennedy, King, and Vespignani}{Lazer
		et~al\mbox{.}}{2014}]%
	{lazer2014parable}
	\bibfield{author}{\bibinfo{person}{David Lazer}, \bibinfo{person}{Ryan
			Kennedy}, \bibinfo{person}{Gary King}, {and} \bibinfo{person}{Alessandro
			Vespignani}.} \bibinfo{year}{2014}\natexlab{}.
	\newblock \showarticletitle{The parable of Google Flu: traps in big data
		analysis}.
	\newblock \bibinfo{journal}{{\em Science\/}} \bibinfo{volume}{343},
	\bibinfo{number}{14 March} (\bibinfo{year}{2014}).
	\newblock
	
	
	\bibitem[\protect\citeauthoryear{Santillana, Nguyen, Dredze, Paul, Nsoesie, and
		Brownstein}{Santillana et~al\mbox{.}}{2015}]%
	{santillana2015combining}
	\bibfield{author}{\bibinfo{person}{Mauricio Santillana},
		\bibinfo{person}{Andr{\'e}~T Nguyen}, \bibinfo{person}{Mark Dredze},
		\bibinfo{person}{Michael~J Paul}, \bibinfo{person}{Elaine~O Nsoesie}, {and}
		\bibinfo{person}{John~S Brownstein}.} \bibinfo{year}{2015}\natexlab{}.
	\newblock \showarticletitle{Combining search, social media, and traditional
		data sources to improve influenza surveillance}.
	\newblock \bibinfo{journal}{{\em PLoS Comput Biol\/}} \bibinfo{volume}{11},
	\bibinfo{number}{10} (\bibinfo{year}{2015}), \bibinfo{pages}{e1004513}.
	\newblock
	
	
	\bibitem[\protect\citeauthoryear{Serfling}{Serfling}{1963}]%
	{serfling1963methods}
	\bibfield{author}{\bibinfo{person}{Robert~E Serfling}.}
	\bibinfo{year}{1963}\natexlab{}.
	\newblock \showarticletitle{Methods for current statistical analysis of excess
		pneumonia-influenza deaths}.
	\newblock \bibinfo{journal}{{\em Public health reports\/}}
	\bibinfo{volume}{78}, \bibinfo{number}{6} (\bibinfo{year}{1963}),
	\bibinfo{pages}{494}.
	\newblock
	
	
	\bibitem[\protect\citeauthoryear{Yang, Santillana, and Kou}{Yang
		et~al\mbox{.}}{2015}]%
	{yang2015accurate}
	\bibfield{author}{\bibinfo{person}{Shihao Yang}, \bibinfo{person}{Mauricio
			Santillana}, {and} \bibinfo{person}{SC Kou}.}
	\bibinfo{year}{2015}\natexlab{}.
	\newblock \showarticletitle{Accurate estimation of influenza epidemics using
		Google search data via ARGO}.
	\newblock \bibinfo{journal}{{\em Proceedings of the National Academy of
			Sciences\/}} \bibinfo{volume}{112}, \bibinfo{number}{47}
	(\bibinfo{year}{2015}), \bibinfo{pages}{14473--14478}.
	\newblock
	
}	
\end{thebibliography}

\end{document}